\documentclass[conference]{IEEEtran}

\usepackage{cite}
\usepackage{amsmath,amssymb,amsfonts}
\usepackage{algorithmic}

\usepackage{textcomp}

\usepackage{tikz}
\usepackage{xcolor}
\usepackage{graphicx}

\usepackage{url}
\usepackage{hyperref}
\usepackage{cleveref}
\hypersetup{pdfborder=0 0 0}

\def\BibTeX{{\rm B\kern-.05em{\sc i\kern-.025em b}\kern-.08em
    T\kern-.1667em\lower.7ex\hbox{E}\kern-.125emX}}
\begin{document}

\newcommand \copyrighttext {
  \footnotesize \textcopyright 2024 IEEE. Personal use of this material is permitted. Permission from IEEE must be obtained for all other uses, in any current or future media, including reprinting/republishing this material for advertising or promotional purposes, creating new collective works, for resale or redistribution to servers or
  lists, or reuse of any copyrighted component of this work in other works. DOI: \href{https://doi.org/10.1109/RE59067.2024.00047}{10.1109/RE59067.2024.00047}.
}

\newcommand\copyrightnotice{
    \begin{tikzpicture}[remember picture,overlay]
    \node[anchor=south,yshift=10pt] at (current page.south) {\fbox{\parbox{\dimexpr\textwidth-\fboxsep-\fboxrule\relax}{\copyrighttext}}};
    \end{tikzpicture}
}

\title{Measuring the Fitness-for-Purpose of Requirements: An initial Model of Activities and Attributes}

\author{
    \IEEEauthorblockN{
        1\textsuperscript{st} Julian Frattini, 
        3\textsuperscript{rd} Davide Fucci, \\
        4\textsuperscript{th} Michael Unterkalmsteiner, 
        5\textsuperscript{th} Daniel Mendez*}
    \IEEEauthorblockA{\textit{Blekinge Institute of Technology}\\
        Karlskrona, Sweden \\
        \{firstname\}.\{lastname\}@bth.se}
    \and
    \IEEEauthorblockN{
        2\textsuperscript{nd} Jannik Fischbach}
    \IEEEauthorblockA{\textit{Netlight Consulting GmbH and *fortiss GmbH}\\
        Munich, Germany \\
        jannik.fischbach@netlight.com}
}

\maketitle

\copyrightnotice

\begin{abstract}
    Requirements engineering aims to fulfill a purpose, i.e., inform subsequent software development activities about stakeholders' needs and constraints that must be met by the system under development.
    The quality of requirements artifacts and processes is determined by how fit for this purpose they are, i.e., how they impact activities affected by them.
    However, research on requirements quality lacks a comprehensive overview of these activities and how to measure them.
    In this paper, we specify the research endeavor addressing this gap and propose an initial model of requirements-affected activities and their attributes.
    We construct a model from three distinct data sources, including both literature and empirical data.
    The results yield an initial model containing 24 activities and 16 attributes quantifying these activities.
    Our long-term goal is to develop evidence-based decision support on how to optimize the fitness for purpose of the RE phase to best support the subsequent, affected software development process. 
    We do so by measuring the effect that requirements artifacts and processes have on the attributes of these activities. 
    With the contribution at hand, we invite the research community to critically discuss our research roadmap and support the further evolution of the model.
\end{abstract}

\begin{IEEEkeywords}
    requirements engineering, requirements quality, literature review, interview study, activity
\end{IEEEkeywords}

\section{Introduction}
\label{sec:intro}

Requirements engineering (RE) is a means to an end and aims to fulfill a purpose, i.e., to inform subsequent activities of the software development life cycle about the needs and constraints of relevant stakeholders~\cite{femmer2018requirements}.
Therefore, requirements artifacts and processes must be fit for purpose.
This fitness for purpose is determined by the attributes of the software development activities that are affected by requirements artifacts or processes~\cite{frattini2023requirements}.
For example, a requirements specification is considered fit for purpose when \textit{implementing} (activity) its implied features works \textit{correctly}, \textit{completely}, and \textit{quickly} (attributes), among other attributes.
In that sense, we should judge the quality of requirements (and RE) based on the extent to which they are fit for purpose, i.e., how they impact the attributes of requirements-affected activities~\cite{femmer2015activities}.
Still, research on requirements quality is dominated by studies aiming to determine the quality of a requirements specification solely based on normative metrics~\cite{frattini2022live}.

Recent endeavors to nuance requirements quality research with this activity-based perspective are promising~\cite{femmer2015activities,femmer2018requirements}, but have so far not seen adoption in practice~\cite{frattini2023requirements}.
One reason for this is the lack of an overview of software development activities that are affected by requirements engineering as well as their measurable attributes.
This gap was acknowledged in previous requirements quality research~\cite{frattini2024identifying,femmer2015activities} and is one milestone on requirements quality research roadmaps~\cite{femmer2018requirements,frattini2023requirements}.
The overview of the activities that are potentially affected by RE would offer guidance on which activities determine the fitness for purpose of RE processes and artifacts.
Furthermore, an overview of the activities' attributes would offer guidance on how to measure their performance.
Consequently, we formulate the following research questions:

\begin{itemize}
    \item \textbf{RQ1}: Which software development activities are affected by requirements artifacts?
    \item \textbf{RQ2}: By which attributes are requirements-affected activities evaluated?
\end{itemize}

This paper initializes the endeavor to create and maintain an overview of requirements-affected activities and attributes answering the research questions.
As the first step, we inductively construct an initial model from three distinct data sources (\Cref{sec:method}).
The model contains 24 activities like \textit{implementing}, \textit{testing}, and \textit{estimating effort}, and characterizes them with 16 attributes including \textit{duration}, \textit{completeness}, and \textit{correctness} (\Cref{sec:results}).
The paper further describes how to apply the model in research and practice and how future research will advance the endeavor (\Cref{sec:plan}).
We disclose all material, data, and source code\footnote{Archived at \url{https://zenodo.org/doi/10.5281/zenodo.10869626}} to facilitate this community endeavor.

\section{Background and Related Work}
\label{sec:related}

\subsection{Requirements Use in SE}

\begin{figure*}
    \centering
    \includegraphics[width=\textwidth]{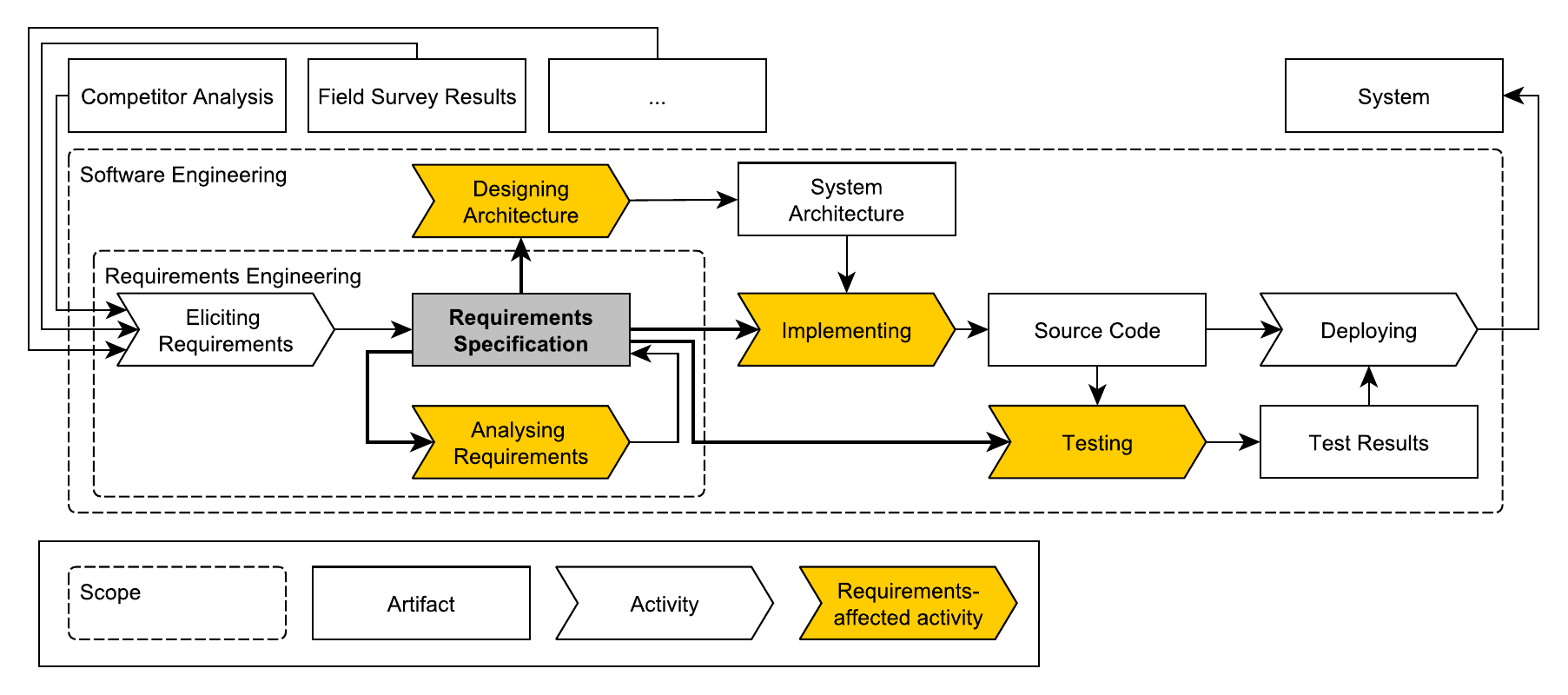}
    \caption{Simplified example of SE-relevant activities}
    \label{fig:activitiesinse}
\end{figure*}

We consider as an \emph{activity} any SE-relevant process performed by a (human or software) agent that uses one or more input artifacts and produces one or more output artifacts~\cite{femmer2015activities}.
\Cref{fig:activitiesinse} visualizes a simplified overview of SE activities, the artifacts they use as an input and produce as an output, and their scope.
For example, the \textit{implementing} activity receives several input artifacts like a requirements specification and system architecture to produce output artifacts like source code.

We consider an activity \textit{requirements-affected} if at least one of its input artifacts is a requirements artifact (yellow activities in \Cref{fig:activitiesinse}).
The aforementioned implementing activity is requirements-affected because it considers a requirements specification as an input.
In the simplified example in \Cref{fig:activitiesinse}, the requirements elicitation and the deployment activity are not requirements-affected.
It is, however, possible that the requirements elicitation activity may be affected by requirements artifacts of previous projects and sprints or that explicit deployment requirements exist.

\subsection{Requirements Quality}

Since requirements artifacts are used as input to requirements-affected activities, the artifacts' quality affects the quality of these activities and their output~\cite{wagner2019status}.
For example, a vague requirements specification may lead to incorrect or missing features and reduced customer acceptance~\cite{fernandez2017naming}.
These quality defects are more expensive to fix the later they are addressed~\cite{boehm1984software}:
Revising a vague requirements specification is less expensive than redeveloping a faulty system built on it.
Therefore, organizations aim to detect and remove requirements quality defects as early as possible~\cite{montgomery2022empirical}.

However, requirements quality research focuses predominantly on normative quality factors~\cite{frattini2022live} that do not consider an impact on affected activities~\cite{montgomery2022empirical,frattini2023requirements}.
For example, the use of \textit{passive voice} is often advised against in literature~\cite{kof2007treatment,genova2013framework,pohl2016requirements} despite a lack of empirical evidence for its negative consequences~\cite{krisch2015myth,frattini2024second,femmer2014impact}.
This fosters skepticism of organizations to adopt requirements quality research~\cite{femmer2018scout,franch2020practitioners}.

To address this issue, Femmer et al. proposed the perspective of \textit{activity-based} requirements quality~\cite{femmer2015activities}.
This perspective entails that requirements are only as good as they support the activities in which they are used~\cite{femmer2018requirements}, i.e., requirements quality depends on the performance of requirements-affected activities.
Specifying requirements quality as fitness-for-purpose to support affected activities necessitates requirements quality research to understand requirements-affected activities, i.e., it requires identifying and measuring activities affected by a requirements artifact~\cite{frattini2023requirements}.

Without a systematic elicitation of requirements-affected activities prior to investigating the quality of a requirements artifact, researchers risk drawing incomplete conclusions.
For example, Ricca et al. investigate the effect of screen mock-ups on requirements comprehension~\cite{ricca2010effectiveness} and conclude that providing screen mock-ups improves the understandability of requirements.
Femmer et al. confirm this effect but contrast that they simultaneously have a negative effect on requirements maintainability~\cite{femmer2015activities}.
Systematic studies on activity-based requirements quality agree that an overview of requirements-affected activities and their attributes is necessary to advance relevant requirements quality research~\cite{femmer2015activities,femmer2018requirements,frattini2023requirements}.

\subsection{Related work}
\label{sec:background:related}

Requirements engineering literature contains several studies about the impact of requirements quality on subsequent software development activities.
For example, Kamata et al.~\cite{kamata2007does} and Zowghi et al.~\cite{zowghi2002study} empirically investigated the impact of requirements quality on project success measured in time and cost overrun.
Similarly, Knauss et al. studied the impact of requirements quality on project success measured by customer satisfaction~\cite{knauss2009investigating}.
These studies generalize the affected activities and summarize their effect on the overall project outcome.

Studies focusing on more specific activities include Chari et al. investigating the impact of requirements defects on injected software defects~\cite{chari2018impact}, and Femmer et al. relating the use of passive voice to the domain modeling activity~\cite{femmer2014impact}.
On the other hand, some studies expand the scope of affected activities.
Damian et al. conducted a longitudinal case study observing a full project development lifespan and measured the tradeoffs of a revised RE process on several activities like communication, effort estimations, and implementation~\cite{damian2006empirical}.
Mendez et al. conducted a large-scale, global survey of perceived problems in RE and their effects on activities, including designing, implementing, and organizing~\cite{fernandez2017naming}.

Research on traceability between software development artifacts constitutes another closely related domain.
Several secondary studies have summarized traceability research and identified artifacts that are commonly connected~\cite{borg2014recovering,charalampidou2021empirical}.
Although requirements artifacts are prominent targets of trace links, they are typically connected to other artifact types, not the activities that produce them~\cite{borg2014recovering}.
These artifact types can be used to infer the producing activities, though the inferred activities typically remain on a very high level~\cite{charalampidou2021empirical}.
Furthermore, this limitation excludes by design all activities that do not necessarily or only rarely produce artifacts, like, for example, informal reviewing, modifying existing artifacts, assessing feasibility, or estimating effort.

In summary, none of these previously mentioned primary studies systematize the affected activities and their attributes but rather select the studied impact based on the availability of data or anecdotal hypotheses, and traceability research exhibits significant limitations regarding the identification of these activities.
Only two studies known to the authors attempt to explicate the affected activities.
Femmer et al. elicited the activities affected by specific requirements artifacts at a case company and determined the qualitative impact of requirements defects on them~\cite{femmer2015activities}.
In a similar study, Frattini investigated requirements quality factors relevant to a case company and their impact on subsequent activities~\cite{frattini2024identifying}.
Both studies prototype a model of requirements-affected activities for the specific context but acknowledge the need for a more systematic and comprehensive overview.

\section{Goal and Early Method}
\label{sec:method}


One goal of activity-based requirements quality research is to create and maintain a comprehensive model of requirements-affected activities and their attributes exhibiting the following properties~\cite{femmer2018requirements,frattini2023requirements}:

\begin{enumerate}
    \item \textbf{Applicability}: The model can represent all requirements-affected activities and attributes in any given SE context.
    \item \textbf{Suitability}: The model can be used to evaluate relevant activities by means of their attributes.
    \item \textbf{Extensibility}: The model can be extended with new activities or attributes.
    \item \textbf{Usability}: The model can be accessed and comprehended by software engineers.
\end{enumerate}


In this study, we contribute the first version of this model.
Since we are not aware of any systematic prior work collecting requirements-affected activities and their attributes~\cite{femmer2018requirements,frattini2023requirements}, we surveyed different data sources for textual descriptions of SE activities that use requirements artifacts as input.
From these textual mentions, we inductively construct a model of requirements-affected activities and their attributes by employing thematic synthesis as proposed by Cruzes and Dyb\aa ~\cite{cruzes2011recommended}.

\subsection{Data Collection}

To ensure the property of applicability as mentioned above, we collected data from three distinct sources described in the following three subsections:
a systematic review of experimentation literature (\Cref{sec:method:collection:slr}), an interview study (\Cref{sec:method:collection:interview}), and a literature study on software process models (\Cref{sec:method:collection:spm}).

\subsubsection{Systematic Literature Review}
\label{sec:method:collection:slr}

The first source of textual descriptions of requirements-affected activities and their attributes that we considered were scientific studies reporting controlled experiments in which the experimental task involves human subjects and considers requirements as an input artifact.
These experimental tasks simulate requirements-affected SE activities performed by practitioners. 
The dependent variables in these experiments are eligible attributes describing the performance of the activity.
We adopted the systematic literature survey method employed by Sj{\o}berg et al.~\cite{sjoberg2005survey}.

\textbf{Database selection.}
To ensure that our database search for eligible primary studies targets publications relevant to SE we pre-selected eligible journals and conferences (from hereon out collectively called \textit{venues}) from the CORE ranking\footnote{\url{https://www.core.edu.au/}} whose field of research is software engineering.
To ensure a high quality of the primary studies, we constrained the venues to those of rank A* or A.
A few select venues of lower rank that are particularly relevant to the topic constituted an exception.
These included the \textit{Requirements Engineering Journal}, the \textit{Journal of Software: Evolution and Process}, the \textit{International Working Conference on Requirements Engineering: Foundation for Software Quality}, the \textit{International Conference on Product-Focused Software Process Improvement}, and the \textit{Euromicro Conference on Software Engineering and Advanced Applications}, which all have a core rank of B.
Additionally, we removed all venues that host computer science rather than SE topics. 
This task was performed by three authors in conjunction to ensure reliability.
The final database selection contained 35 venues (10 journals and 25 conferences).

\textbf{Database search.}
We performed a keyword-based database search for each included venue with the keywords \textit{experiment*} as well as \textit{requirement*} (or the synonyms \textit{srs} or \textit{specification*}).
These keywords limited the retrieved primary studies to those (1) describing an experiment and (2) involving requirements at least to some degree.
We executed the database search via Scopus\footnote{\url{https://www.scopus.com/search/form.uri?display=advanced}} and in four cases, where Scopus did not index publications of that venue, via the ACM Digital Library.\footnote{\url{https://dl.acm.org/}}
The search string per venue consisted of the two sets of keywords as well as a limitation to the venue via its title.
For example, the search string for the ACM Computing Surveys journal in Scopus looked as follows: \texttt{SRCTITLE ( computing AND surveys ) AND TITLE-ABS-KEY ( requirement* OR srs OR specification* ) AND TITLE-ABS-KEY ( experiment* )}.
The search per venue returned between 1 (e.g., from the \textit{European Conference on Object-Oriented Programming}) and 175 (from the \textit{Journal of Systems and Software}) primary studies for a total of 1446 studies.

\textbf{Inclusion.}
Next, we performed an inclusion phase to ensure the following properties of primary studies expressed by the two inclusion (I1 and I2) and four exclusion criteria (E1-E4):

\begin{itemize}
    \item I1: The primary study presents an experiment with human subjects as one of its core contributions.
    \item I2: The experimental task uses a requirements specification as an input.
    \item E1: The experimental task is a requirements review.
    \item E2: The study is not written in English.
    \item E3: The publication is not available via the university's access program.
    \item E4: The study is a duplicate of or extended by an already included study.
\end{itemize}

I1 ensures that eligible primary studies present a proper experiment (regardless of whether it is controlled or quasi) that involves human subjects.
Otherwise, the experimental task would not simulate an SE activity, the concept of interest.
This excludes, for example, experiments in which machine learning algorithms of different configurations are compared on a classification task.
I2 ensures that the activity is requirements-affected.
E1 explicitly excludes requirements review tasks, i.e., requirements defect detection and removal activities.
The purpose of identifying requirements-affected activities is to optimize the affecting requirements in a way that improves their impact on the activities.
This optimization process is the requirements review.
Hence, we excluded these studies to avoid a circular impact, i.e., suggesting to optimize requirements for the reviewing activity, which is exactly this optimization.
E2 and E3 exclude inaccessible studies, and E4 removes content duplicates.
Primary studies were considered for the next data analysis step when they met all two inclusion and none of the exclusion criteria.
The first author conducted the inclusion step based on the title, abstract, and keywords.
Out of 1446 primary studies, 145 (10.3\%) were included.
To ensure the reliability of this subjective process, the second author independently performed the inclusion task on 75 (i.e., 5.2\%) randomly selected studies.
We calculate the inter-rater agreement using Bennett's S-Score~\cite{bennett1954communications}, which is robust against uneven marginal distributions~\cite{feng2015mistakes}.
The inter-rater agreement yields a value of 92\%, which we deem sufficient to instill confidence in this subjective task.

\textbf{Data Extraction.}
The first author reviewed all 145 included primary studies and extracted, for each human-subject experiment in each study, (1) the description of the experimental task and (2) all dependent variables measured to evaluate the performance of the task.
The description of the experimental task constituted the source of requirements-affected activities, and the dependent variables were the source of their attributes.
While reviewing the full text of the studies, 22 studies revealed to not, in fact, meet all inclusion criteria other than the title, abstract, and keyword had suggested.
We excluded these 22 studies from further processing.

Additionally, we excluded extractions where the attribute description did not \textit{imply a valuation}.
Because our goal was to identify attributes that quantify the \textit{performance} of their respective activity, eligible attributes must be \textit{valuating}---i.e., values of that attribute must imply a degree of performance.
While attributes do not necessarily have to be measured on an interval scale (i.e., it is not important to associate an interval of the attribute, like a certain amount of minutes for the attribute \textit{duration}, with a specific level of quality), it has to be at least on an ordinal scale---i.e., the sign of the interval is important (more minutes of duration is bad, less minutes of duration is good).
For example, if the dependent variable of an experiment investigating the activity of \textit{estimating effort} is the \textit{estimated amount of hours}~\cite{molokken2005expert}, then this data point(i.e., pair of activity and attribute)  was excluded as a higher or lower value of that attribute does not automatically make it good or bad due to the lack of ground truth.
If, instead, the dependent variable was \textit{precision}, i.e., how close the estimated amount of hours is to actual implementation time, then the data point would be included as a higher value of precision (i.e., an estimation that is closer to the actual time) is better.
This process eliminated 12 descriptions of non-valuating attributes.
To assess the validity of this process, the third author independently repeated the task on a sample of 12 data points, which consisted of 6 random samples from each of the two classes (valuation vs. no valuation), and we measured the inter-rater agreement using Cohen's Kappa~\cite{cohen1960coefficient} since the classes have an even marginal distribution~\cite{feng2015mistakes}.
The first overlap achieved a Cohen's Kappa of only 33.3\%, which emphasized the complexity of the task.
The two authors reconvened, discussed the differences, reformulated the exclusion criteria, and repeated the labeling.
The second overlap achieved a score of 83.3\%, which represents a sufficient reliability of the step.

The extraction produced 142 descriptions of experimental tasks and 355 descriptions of dependent variables.
Several experimental tasks were evaluated via multiple dependent variables, which is why the 355 resulting data points contain repeated descriptions of experimental tasks.

\subsubsection{Interview Study}
\label{sec:method:collection:interview}

The second source of textual descriptions of requirements-affected activities and their attributes that we consider were reports from industry practitioners about the usage of requirements specifications in subsequent SE activities.
To this end, we evaluated the transcripts of a previously conducted interview study~\cite{frattini2024identifying}.

\textbf{Interview Participants.}
The first author conducted the interview study in a large, globally distributed software development organization that specifies requirements using both free-form and constrained natural language (use cases) prior to each development cycle.
A contact at the organization provided a sample of eight software engineers directly responsible for processing requirements specifications and developing solution specifications based on them.
These eight engineers represent the majority of personnel in their role in the team that was involved in the study.
The interview participants had an average of 3.5 years of experience in their role, 7.5 years with the organization, and 15.3 years as software engineers.

\textbf{Interview.}
The original purpose of the interview was to identify which quality defects practitioners perceive in the requirements specifications that they process~\cite{frattini2024identifying}.
Because the elicitation of quality defects entailed mentioning what kind of subsequent activity is affected by this defect, the generated data served to identify requirements-affected activities and their attributes.
For example, stating that vague requirements lead to a delay of the testing phase contains the requirements-affected \textit{testing} activity and its attribute \textit{duration}.
To guide the semi-structured interview, we developed a protocol.
The protocol contained, among demographic questions, one prompt per type of requirements quality.
The types of requirements quality were derived from Montgomery et al.~\cite{montgomery2022empirical} and covered, among others, ambiguity, completeness, and traceability.

\textbf{Data Extraction.}
All eight one-hour-long interviews were recorded, automatically transcribed using a speech-to-text conversion tool,\footnote{\url{https://www.descript.com/}}, and verified by the first author.
Then, the first author extracted from the transcripts each mention of an activity affected by a requirements quality defect and how this effect was measured.
The extraction produced 55 descriptions of affected activities but no descriptions of how this effect was measured on them.

\subsubsection{Literature Study}
\label{sec:method:collection:spm}

The third source of textual descriptions of requirements-affected activities and their attributes that we consider were descriptions of software process models.
Software process literature describes processes and products of the SE life cycle and, hence, contains information about which activities are affected by requirements.
Since software process literature is fairly mature~\cite{sommerville1996software}, we have access to reliable summaries of process models.

\textbf{Literature.}
We selected the book ``Software Process Definition and Management'' by Münch et al.~\cite{munch2012software} as a reliable summary of software process literature.
The first author reviewed the descriptions of all seven lifecycle models, which cover the waterfall model~\cite{royce1987managing}, iterative enhancement~\cite{boehm1984software}, prototyping, the spiral model~\cite{boehm1988spiral}, the incremental commitment spiral model~\cite{boehm2008guide}, Unified Process~\cite{jacobson1999unified}, and Cleanroom Development~\cite{mills1987cleanroom}.
The first author extracted all textual mentions of requirements-affected activities and their attributes as prescribed by the lifecycle model.
This extraction produced 21 textual descriptions of activities and one explicit description of an attribute.

\subsection{Data Analysis}

\textbf{Coding.}
The data collection phase over the three sources culminated in a table containing 218 textual descriptions of requirements-affected activities and 356 textual descriptions of their attributes.
In the absence of a prior theory or model of requirements-affected activities, we resorted to an inductive coding process~\cite{cruzes2011recommended}.
The first and third authors jointly established the level of granularity of the codes that were applied to the textual descriptions and documented this process in a guideline.
The first author then performed the coding process independently and, upon completion, verified the assigned codes with the third author.
For each pair of textual descriptions of an activity and attribute, we coded four concepts:

\begin{enumerate}
    \item Activity: the requirements-affected activity
    \item Activity attribute: a property evaluating an activity
    \item Artifact: an output artifact produced by the activity
    \item Artifact attribute: a property evaluating an artifact
\end{enumerate}

The distinction of artifacts from activities was necessary since some activities were not evaluated directly but rather by the artifacts they produced.
For example, \textit{duration} is an attribute of the \textit{implementing} activity, but several studies additionally evaluate that activity by measuring the \textit{coupling} (artifact attribute) of the resulting \textit{source code} (artifact).

\textbf{Consolidation.}
The inductive coding process produced 24 unique codes for activities, 16 for activity attributes, 21 for artifacts, and 26 for artifact attributes.
The first and third authors then created an abstraction hierarchy of identified activities and artifacts based on the guide to the software engineering body of knowledge~\cite{bourque2004swebok}.
For example, both the \textit{planning} and the \textit{estimating effort} activities are sub-types of the more abstract \textit{managing} activity~\cite{bourque2004swebok}.
We decided to merge the activities \textit{interpreting} and \textit{understanding} with \textit{comprehending} as none of the data sources sufficiently distinguished between them.
Future studies differentiating them properly are necessary.

Once the hierarchy emerged, we associated each activity and artifact with the respective attributes that our data sources reported to characterize them.
Whenever all activities or artifacts of a hierarchical group shared an attribute, we moved it to the higher-level activity or artifact for conciseness.
Additionally, we made educated assumptions about the transferability of some attributes.
For example, even though our data did not contain an instance of \textit{duration} being evaluated on every activity, it is safe to assume that every activity can be characterized and evaluated by its duration. 
This step introduces slight subjectivity but improves the applicability of the model.

\subsection{Data Availability}

To achieve the goals of usability and extensibility of the resulting model, we disseminate it via GitHub.\footnote{Available at \url{https://github.com/JulianFrattini/gere-r3a}}
The repository contains a reference to all considered data sources, guidelines and protocols for the data extraction, and a specification of the current model of requirements-affected activities and their attributes.
More importantly, it contains guidelines on how to contribute new or revise existing activities and attributes.
Using the version control system of GitHub\footnote{\url{https://docs.github.com/en/get-started/using-git/about-git}} we will foster a collaborative evolution of the model.

\section{Results}
\label{sec:results}

\subsection{Requirements-affected Activities and their Attributes}
\label{sec:results:model}

\begin{figure*}
    \centering
    \includegraphics[width=\textwidth]{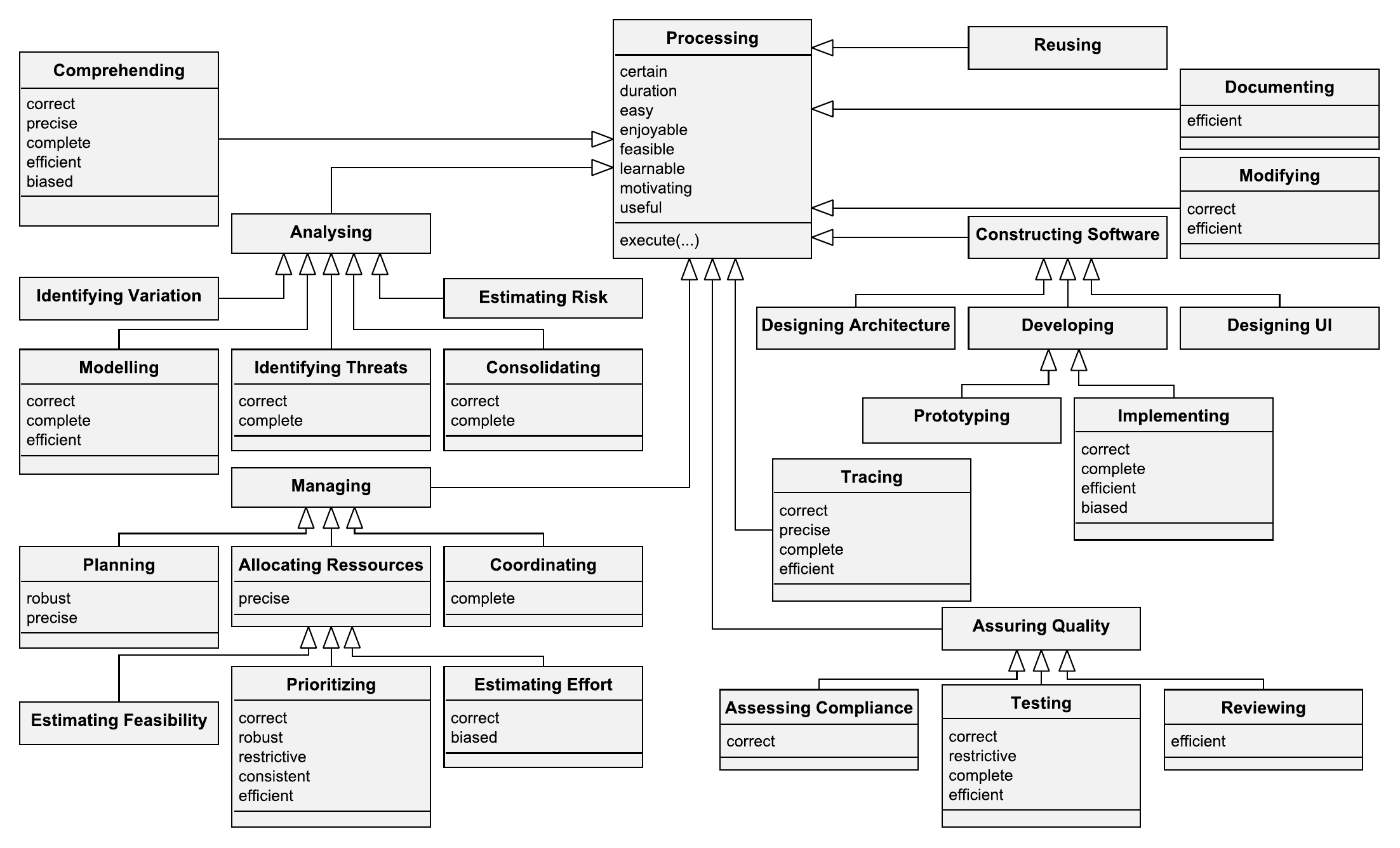}
    \caption{Model of requirements-affected activities and their attributes}
    \label{fig:r3a}
    \vspace{-.4cm}
\end{figure*}

\Cref{fig:r3a} visualizes the initial model of requirements-affected activities and their attributes.
The model is structured like a UML class diagram and makes use of the inheritance relationship.
An activity, represented as a UML class, that inherits from another activity also exhibits its attributes.
For brevity, artifacts are excluded from the visualization. 
The replication package contains an extended model that includes the artifacts.
The root of the inheritance tree is the abstract activity \textit{processing}, which represents every executable activity.
The model contains several activities that are commonly considered in research as requirements-affected activities, like modeling, prioritizing, implementing, and testing.
Another prominent spot is taken by the merged activity \textit{comprehending}, which dominates the distribution of activities among both experimental literature and interview statements.
This correlates to the prominence of ambiguity among the attributes of requirements quality in empirical research~\cite{montgomery2022empirical} and is supported by the fact that this activity precedes every other activity~\cite{femmer2015activities}.
The model, furthermore, contains several less commonly investigated activities.
For example, Murakami et al. investigate the activity of \textit{code review} in which subjects are provided with a requirements specification~\cite{murakami2017wap}.
\textit{Consolidating} larger sets of requirements to identify a semantically equivalent subset~\cite{natt2006experiment,wnuk2012replication} is another rare example.
The model also contains activities that did not appear in experimental studies but were reported by interview participants or prescribed by software process literature.
The activity of \textit{prototyping} is such an example that was both mentioned during the interviews and as part of lifecycle models.
Furthermore, the following activities were all named by interview participants but not considered in the experimentation literature:
\textit{coordinating} internal stakeholders based on a requirement, \textit{reusing} artifacts like source code based on a new requirement, and \textit{estimating feasibility} of a requirement.
The attributes recorded in the model also show a varying distribution of prevalence.
The most commonly encountered attributes of an activity are \textit{duration}, \textit{correctness}, and \textit{completeness}.
These represent both simple-to-measure and critical properties of most activities.
Additionally, we observed several attributes related to the effect that the activity has on the executing agent, for example, how \textit{certain} an agent feels when executing the activity, how \textit{easy}, \textit{enjoyable}, \textit{motivating}, and \textit{useful} they perceive it to be, and how \textit{learnable} the activity was.
Rarely mentioned attributes include how \textit{robust} an activity is against errors and how \textit{biased} an activity becomes given some controlled stimulation.

\subsection{Implications}
\label{sec:results:implications}

\subsubsection{Implications for Research}

The results contain multiple implications for requirements engineering and, specifically, requirements quality research.
Firstly, the distribution of activities and attributes among the three data sources hints at potential research gaps. 
For example, the above mentioned activities of prototyping, coordinating, reusing, and estimating have not appeared in the sample of primary studies.
Secondly, the model provides guidance for comprehensive measurements of the software development life cycle with respect to the impact of requirements artifacts and processes.
As determined by Femmer et al., only a holistic view of all requirements-affected activities will reliably determine the impact of any treatment in requirements artifacts or processes~\cite{femmer2015activities}.
This affects all comparative studies in requirements engineering, i.e., all controlled and quasi-experiments aiming to evaluate the impact of a quality defect or the benefit of a new method.
Only by measuring this impact on all requirements-affected activities in terms of their attributes and summarizing the total benefit or drawback, a holistic decision on the benefit or harm of any treatment can be made.
While we certainly do not suggest that any comparative study from here on out must necessarily consider all 24 activities simultaneously, the model of requirements-affected activities provides at least a framework that allows integrating the results of multiple studies investigating the effect of the same treatment on different activities to one, overall conclusion.


\subsubsection{Implications for Practice}

The resulting initial model of requirements-affected activities and their attributes may serve practitioners as an overview of activities to measure when attempting to understand the fitness for purpose of their requirements.
The model emphasizes the diversity of activities that may be affected by requirements but also the diversity of metrics by which they can be evaluated.
While attributes like completeness, correctness, and duration are likely to be covered in key performance indicators of organizations, attributes like usefulness, ease of use, and learnability may often be neglected.
Further practical use of the model for quantitative comparisons requires future work and will be discussed in \Cref{sec:plan}.

\subsection{Limitations}
\label{sec:results:limitations}

This study exhibits the following limitations. 
Firstly, the data extraction phase was only performed by one researcher. 
This introduces the possible risk that relevant information from the bodies of text was missing from the textual descriptions that were later coded.
Secondly, the interview study was not performed with the research questions stated in \Cref{sec:intro} in mind.
Instead, the main theme of the interview study was centered around the broader scope of requirements quality~\cite{frattini2024identifying}.
However, confirming previous studies that proposed that requirements quality inevitably depends on requirements-affected activities~\cite{femmer2015activities,femmer2018requirements,frattini2023requirements}, the responses of interview participants naturally contained information that contributed to answering our research questions.
Hence, we deem the interview data as an eligible data source for this study.
Thirdly, every step of the study where we depart from purely summarizing and reporting data and instead interpret it introduces researchers' bias.
This is particularly evident in the conscious merging of the understanding, interpreting, and comprehending activity but also in the assumption about the transferability of several activities' attributes.
This step was necessary to elevate the model beyond a systematic summary toward an evaluation framework as demanded in previous research roadmaps~\cite{femmer2018requirements,frattini2023requirements}.
We documented all interpretative steps and disclosed them in our replication package to allow other researchers to scrutinize these decisions.
Finally, we address the threat to external validity.
Full generalizability was out of the scope of the goals of this study, but we, nevertheless, briefly discuss all threats to external validity in order to justify the research plan as presented in \Cref{sec:plan}.
One threat to the generalizability stems from the sampling of the literature survey, which only considers a specific set of SE-relevant venues and categorically excludes workshops.
Additionally, the literature review is limited to experiments and excludes other methods like case studies.
Another threat stems from the sample of interview participants, which represent only one team of only one company.

\section{Research Plan}
\label{sec:plan}

\subsection{Model Extension}
\label{sec:plan:extension}

The limitations mentioned in \Cref{sec:results:limitations} necessitate the extension of the model to achieve goals 1 (applicability) and 2 (suitability) stated in \Cref{sec:method}.
Both the applicability and the suitability are inhibited by the potential incompleteness of the model.
Hence, we plan to repeat the early method presented in \Cref{sec:method}.
Two immediately planned extensions are (1) repeating the systematic literature survey on workshop papers and (2) replicating the interview study in different companies and teams.
Because of the extensive documentation of data collection methods for both empirical data (i.e., interview transcripts) and meta-research (i.e., primary studies), as well as the data analysis protocol, we anticipate that the model extension can be distributed well within our network of researchers interested in requirements-affected activities.

\subsection{Model Maintenance}
\label{sec:plan:maintenance}

Goals 3 (extensibility) and 4 (usability) stated in \Cref{sec:method} are fulfilled by the design of the chosen dissemination strategy.
The authors of this study will maintain the GitHub repository containing the current content and structure of the model.

\subsection{Model Validation}
\label{sec:plan:validation}

The most significant step of future work is to validate whether the model achieves the four goals stated in \Cref{sec:method}.

\textbf{Validating applicability.}
To test whether the model can represent all requirements-affected activities and attributes in any given SE context, we plan to conduct multiple case studies in different company contexts.
Once the model is deemed sufficiently extensive, we trace requirements artifacts in each case company to every instance of reuse.
The process of tracing requirements artifacts to activities using these artifacts as input shall happen both directly, i.e., by interviewing involved stakeholders, but also indirectly, i.e., by observing which stakeholder accesses the artifact and then following up on the purpose.
The latter accounts for requirements-affected activities that stakeholders are not actively aware of, i.e., in case they unconsciously retrieve information to execute an activity without considering that this makes the activity requirements-affected.
We constitute that the model achieves goal 1 if we do not encounter any requirements-affected activity that has no semantic equivalent in the model.

\textbf{Validating suitability.}
To test whether the model can be used to evaluate relevant activities by means of their attributes, we plan to conduct an empirical study involving all surveyed case companies.
Given the already detected requirements-affected activities, we evaluate these via the attributes associated with the activities in our model to quantify their performance.
We aim to produce two types of empirical investigations from this data.
Firstly, we aim to survey the activities and generate an overview of attribute values for all affected activities.
This overview provides an absolute comparison of the activities and answers questions like ``Which activity phase takes the longest time'' or ``Which development activity is perceived as the least enjoyable?''.
Secondly, we aim to conduct quasi-experiments at the case companies investigating whether certain properties of requirements artifacts or properties have an impact.
For example, the subject of the experiments could be the comparison between two types of template systems for requirements specification~\cite{grosser2023comparative} or the avoidance of specific linguistic structures like passive voice~\cite{frattini2024identifying}.
The subject of the experiments will be aligned with current questions and endeavors of the case companies to optimize their requirements engineering artifacts or process in an evidence-based manner.
The results of the quasi-experiments will be measured in terms of differences in attribute values of all affected activities.
This overview will provide companies with a summary of the effect that the proposed change has on all affected activities.
We constitute that the model achieves goal 2 if the results generated by the surveys and quasi-experiments are accepted by the respective case companies.

\textbf{Validating extensibility.}
To test whether the model can extended with new activities or attributes, aim to involve additional researchers in the model extension presented in \Cref{sec:plan:extension}.
By distributing the task beyond the authors of this study, we determine how easily other researchers can extend the model.
We constitute that the model achieves goal 3 if external researchers extend the model successfully.

\textbf{Validating usability.}
To test whether the model can be accessed and comprehended by software engineers, we plan to facilitate external replications of the validation of goals 1 and 2.
This not only validates whether the model achieves goal 4 but also extends the empirical evidence about the impact of requirements on affected activities in different company contexts.
We constitute that the model achieves goal 4 if external researchers successfully replicate the empirical studies.

\section{Conclusion}
\label{sec:conclusion}

Requirements artifacts and processes fulfill a specific purpose in the software development lifecycle, that is, to inform subsequent activities about the needs and constraints imposed by stakeholders on the system under development~\cite{femmer2015activities}.
How fit requirements artifacts and processes are to fulfill their purpose, i.e., how well they benefit these requirements-affected activities, can be effectively determined when (1) all affected activities are known and (2) the performance of these activities can be evaluated.
The need for a systematic overview of (1) requirements-affected activities as well as (2) the attributes which quantify their performance has been well recognized in requirements quality literature~\cite{femmer2015activities,frattini2024identifying} and evoked the call for a comprehensive model~\cite{femmer2018requirements,frattini2023requirements}.

We answer this call by proposing an initial model of requirements-affected activities and their attributes systematically derived from three distinct data sources.
The model aims to support both researchers by guiding empirical studies concerning the impact of requirements artifacts and processes but also practitioners by offering an overview of attributes that may serve as key performance indicators of their requirements-affected activities.
We envision that this model will be extended and evolved by the requirements engineering community to provide an applicable and suitable model for the task.
We will actively maintain the presented resources to enable and foster this community endeavor.

\section*{Acknowledgment}
This work was supported by the KKS foundation through the S.E.R.T. Research Profile project at Blekinge Institute of Technology.
We further thank Parisa Yousefi and Charlotte Ljungman from Ericsson Karlskrona for facilitating the interview study.

\bibliographystyle{IEEEtran}
\bibliography{material/references}

\end{document}